# Graphene as Transparent Electrode for Direct Observation of Hole Photoemission from Silicon to Oxide

*Rusen Yan(闫汝森),[1,2] Qin Zhang,[1,2] Oleg A. Kirillov,[1] Wei Li,[1,3] James Basham,[1] Alex Boosalis,[1,4] Xuelei Liang,[3] Debdeep Jena,[2] Curt A. Richter,[1] Alan C. Seabaugh,[2] David J. Gundlach,[1] Huili G. Xing[2,]\* and N. V. Nguyen[1,]\**

1. Semiconductor and Dimensional Metrology Division, National Institute of Standards and Technology, Gaithersburg, Maryland 20899, USA

2. Department of Electrical Engineering, University of Notre Dame

Notre Dame, IN 46556, USA

3. Key Laboratory for the Physics and Chemistry of Nano Devices, Peking University

Beijing, China

4. Department of Electrical Engineering and Nebraska Center for Materials and Nanoscience

University of Nebraska-Lincoln, Lincoln, Nebraska 68588, USA

Abstract:

We demonstrate the application of graphene as collector material in internal photoemission (IPE) spectroscopy, which enables direct observation of both electron and hole injections at a Si/Al$_2$O$_3$ interface and overcomes the long-standing difficulty of detecting holes in IPE measurements. The observed electron and hole barrier heights are 3.5 ± 0.1 eV and 4.1 ± 0.1 eV, respectively. Thus the bandgap of Al$_2$O$_3$ can be deduced to be 6.5 ± 0.2 eV, in good agreement with the value obtained by ellipsometry analysis. Our modeling effort reveals that, by using graphene, the carrier injection from the emitter is significantly enhanced and the contribution from the collector electrode is minimal.

The past few years have witnessed rapid growth of interest in graphene due to its promise for use in future electronic and optical devices.[1] Specifically, its high optical transmittance over a wide spectrum (at least 1-6 eV) and electrical conductivity make graphene an attractive candidate as a transparent electrode.[2, 3] These remarkable properties make it an excellent photoexcited carrier collector material for internal photoemission (IPE) spectroscopy. Yan *et al.*[8] and Xu *et al.* recently reported employing graphene as collector to extract material work functions using IPE. Since 1960s, IPE as a measurement science has continually been improved and shown to be a robust technique to characterize the interface properties, charge trapping phenomenon,[4-6] and most commonly, as a quantitative method for determining electronic band alignment.[6, 7] In most of these experiments, a thin metal layer (10-15 nm) is used as an optically semitransparent contact to collect electrons or holes injected from the semiconductor emitter over the energy interfacial barrier formed between the emitter and the insulator. However, a long-standing issue with such a test structure is that, the photocurrent due to hole injection is usually obscured by the unavoidably large electron current from the thin metal contact (collector) over the insulator.[9] Further complicating the experimental observation of hole injection from the semiconductor emitter is the rather limited range over which metal work function can be varied; for most semiconductor systems of interest, the barrier height for holes at the semiconductor-insulator interface is usually higher than the barrier height for electrons at the metal/insulator interface. Consequently, it is difficult to separate the contribution of hole emission and its barrier threshold from the total measured photocurrent.[5, 9] Additionally, semitransparent metals of a practical thickness used in IPE studies have substantial light absorption, particularly in the ultraviolet range that is important in IPE measurements, thus resulting in considerably less incident power absorbed by the semiconductor.[10] The number of photoexcited carriers turns out to be much larger in metal than in the semiconductor, further hindering the detection of holes injected from the semiconductor. Transparent conducting oxides like ITO are not suitable due to their often process-dependent band gaps on the order of 4 eV, which will cut off nearly all the photons with energy larger than that.[11] Goodman et al. attempted to address this experimental challenge by replacing the metal electrode with water.[12] However, this approach is inconvenient since the use of a water electrode significantly complicates the fabrication of devices of interest and the measurement setup.

In this report, we propose and demonstrate an important application of graphene as an elegant solution to this metrology challenge by utilizing graphene as a transparent electrode to collect photo-generated carriers in IPE. Its high transparency over a wide spectral range (IR/Visible/UV) enables direct observation of the hole injection, and facilitates determination of *both* conduction and valence band offsets at the semiconductor-insulator junction or hetero-junction. In addition, hole transition detection will also allow one to characterize other electronic properties at the interface such as interfacial dipoles, carrier trapping effects, hole transport, etc.[9] Finally, with direct and precise measurements of both the electron and hole barrier heights one can accurately deduce the band gap of the insulator, therefore providing a complete energy band diagram alignment of the heterostructure.[5, 10]

We employ a graphene/$Al_2O_3$/Si structure as a technologically important material system to demonstrate the feasibility and utility of our approach. The device structure is schematically depicted in Fig. 1 (a). A 10 nm thick $Al_2O_3$ layer is deposited by atomic layer deposition (ALD) on an RCA-cleaned $p^{++}$-Si substrate. A large-area monolayer graphene sheet grown by chemical vapor deposition (CVD) was transferred onto the prepared $Al_2O_3$/Si substrate.[13] A 100 x 200 µm$^2$ rectangular graphene region was then patterned by oxygen plasma etching. A 180 nm thick

Al contact for probing is deposited on part of the graphene collector to complete the test structure fabrication. The top-view optical image of the finished device is shown in Fig. 1 (b).

The IPE measurement system mainly consists of a 150 W broadband xenon light source and a quarter-meter Czerny Turner, f/4, 1200 line/mm ruled monochromator to provide a spectral range from 1.5 to 5.5 eV. A voltage ($V_{gs}$) is applied between the top Al contact and the $p^{++}$-Si substrate, and a photocurrent ($I_{ph}$) flowing across the graphene/Al$_2$O$_3$/Si heterojunction is recorded by an electrometer as a function of photon energy ($h\upsilon$).

Shown in Fig. 1(c) are the photocurrents, $I_{ph}$, due to either electron or hole transitions between Si (named as gate) and graphene (named as source and grounded) measured as a function of incident photon energy under various gate voltages $V_{gs}$. The oxide flatband voltage, $V_{fb}$, occurs when the net electric field in the oxide, thus the photocurrent, both reach zero for photon energies larger than the barrier threshold. $V_{fb}$ is found to be about 0.6 V with respect to the grounded graphene, which is in good agreement with a previous band alignment analysis.[8] When $V_{gs}$ = -2.9, -2.8, -2.7, -2.6 V, much smaller than $V_{fb}$, the spectral photocurrent tends to go negative for the above threshold photons, which corresponds to the energy diagram depicted in Fig. 1(d). In this case, the electric field in the oxide drives the electrons (photo-excited above the Al$_2$O$_3$ conduction band bottom) from Si into graphene. On the other hand, when $V_{gs}$ = 2.6, 2.7, 2.8, 2.9 V, the reversed electric field drives holes excited in Si into graphene as depicted in the energy diagram in Fig. 1(e). In the latter case, the photo-carriers excited in graphene and injected into Si are negligible since the photon absorption is low for graphene (< 5%) compared to that by Si (> 30%) over the entire spectral range in the measurement. Further evidence of the hole injection will be presented below in the data analysis. This demonstrates that by taking advantage of the uniquely transparent nature of graphene we have overcome the past difficulty of detecting this hole injection in IPE when metals are commonly used as an electrode.

The electron or hole barrier height is directly determined from the photoemission quantum yield (*Y)*, which is obtained from the measured photocurrent, $I_{ph}$, normalized by the incident photon flux.[14] It is well known that the cubic root of the yield near the barrier threshold ($\phi$) is linearly related to photon energy ($h\upsilon$) when the photocurrent is dominated by carriers excited from 3-dimensional semiconductors in an IPE measurement, which follows the equation:[9]

$$Y^{1/3} = A(h\upsilon - \phi) \quad (1)$$

where *A* is a constant dependent on photon intensity. Shown in Fig. 2 (a) and (b) are the $Y^{1/3}$ vs. $h\upsilon$ plots for the negative (electron) and positive (hole) photocurrents, respectively. It can be seen that, the yield starts to increase sharply and linearly near the barrier height threshold. The noticeable features in both injection spectra are the kink at ~ 4.4 eV and the change of slope at ~ 3.5 eV, which is used to differentiate hole injection from electron injection. Generally speaking, the presence of these features offers critical correlations to assessing the origin of photocurrent.[3, 10, 14] The positions of the 3.5 eV and 4.4 eV features in the yield plot align perfectly with the optical singularities ($E_1$ and $E_2$) of crystalline Si, thus indicating that both currents primarily stem from carrier injection (electron or hole) from the Si substrate but not from graphene. The negative bias branches in Fig. 2 (a) due to electron injection from Si, contain both features, which indicates the effective barrier height for electrons ($\phi_e$) is lower than $E_1$, the smaller of the two. The positive bias branches in Fig. 2 (b) due to hole injection from Si show the $E_2$ transition only, and the vanishing of $E_1$ in the yield plot suggests that the barrier height for holes is larger than $E_1$. The transition threshold shifts to a lower energy for higher gate voltages following the well-known Schottky barrier lowering effect.[15] To extract the barrier height at the flatband condition, it is necessary to extrapolate the effective barrier height obtained from the yield plot under non-zero fields to the zero field in the oxide. This field dependence of barrier height can be well-described by the relation:[5, 10]

$$\phi = \phi_0 - q(q/4\pi\varepsilon_0\varepsilon_i)^{1/2}F^{1/2} \tag{2}$$

where $q$ is the fundamental electron charge, $F$ is the oxide field, $\varepsilon_0$ and $\varepsilon_i$ are the vacuum permittivity and the effective permittivity of the oxide, respectively. The zero-field barrier heights of electron ($\phi_e^0$) and hole ($\phi_h^0$) are obtained by a linear fit of $\phi$ versus $F^{1/2}$ as shown in Fig. 3. The barrier height from the top of the Si valence band to the bottom of the $Al_2O_3$ conduction band, $\phi_e^0$, is found to be 3.5 eV ± 0.1 eV; the barrier height from the bottom of the Si conduction band to the top of the $Al_2O_3$ valence band, $\phi_h^0$, is found to be 4.1 eV ± 0.1 eV.

Unlike prior approaches[6, 12, 16, 17] implemented for IPE measurements that suffer from inherent limitations, our approach enables direct observation of the hole transition and provides simultaneous and exclusive information about the conduction and valence band at critical material interfaces. One additional and beneficial outcome from our approach is that the band gap ($E_g$) of the insulator can be easily deduced from the electron and hole energy barrier heights by this simple relation: $E_g^{insulator} = \phi_e^0 + \phi_h^0 - E_g^{semiconductor}$, which can be compared with bandgap values derived from purely optical measurements and modeling. In this particular study using ALD $Al_2O_3$, we find $E_g^{Al2O3} = \phi_e^0$ (3.5 eV) + $\phi_h^0$ (4.1 eV) - $E_g^{Si}$ (1.1 eV) = 6.5 ± 0.2 eV. This method of determining the bandgap can be preferred for some material systems because it is free from possible marring induced by excitonic effects.[9] To verify the bandgap value of our ALD $Al_2O_3$, we also performed vacuum ultraviolet spectroscopic ellipsometry (VUV-SE) measurement on the same $Al_2O_3$/Si structure,[18, 19] revealing a band gap of 6.5 eV ± 0.05 eV, in an excellent agreement with the value determined by IPE. It is worth pointing out that the band gap of $Al_2O_3$ highly depends on growth conditions and the thickness. It is thus expected that the band gap of amorphous $Al_2O_3$ grown by ALD differs from that (~ 9.5 eV) of bulk crystalline $Al_2O_3$.

To investigate further the advantage of using graphene as the collector electrode over traditional metals, we have quantitatively evaluated absorption by each layer in the graphene-oxide-semiconductor (GOS) and metal-oxide-semiconductor (MOS) IPE test structures.[20] Let us consider the case of normal optical incidence in air with a refractive index $n_0 = 1$ into a three-layer stack consisting of semitransparent electrode (metal or graphene), $Al_2O_3$ and Si with a complex and wavelength-dependent refractive index of $n_1$, $n_2$ and $n_3$, respectively.[21] The thickness of the metal or graphene is $d_1$, and that of the oxide is $d_2$. Also assumed is the Si substrate being semi-infinite and $Al_2O_3$ being transparent with a zero imaginary refractive index in the entire optical range. For a single-layer graphene, $n_1$ measured by spectroscopic ellipsometry[22] and a thickness of 0.34 nm are used. With the described geometry, it is straightforward to show the reflection by the entire stack is given by:[23]

$$R = |E_0^-/E_0^+|^2 = \left| \frac{r_1 + r_2 \exp(2\delta_1) + r_3 \exp 2(\delta_1 + \delta_2) + r_1 r_2 r_3 \exp(2\delta_2)}{1 + r_1 r_2 \exp(2\delta_1) + r_1 r_3 \exp 2(\delta_1 + \delta_2) + r_2 r_3 \exp(2\delta_2)} \right|^2 \quad (3)$$

where $r_i$'s are the Fresnel reflection coefficients defined as:

$$r_1 = \frac{n_0 - n_1}{n_0 + n_1}, \quad r_2 = \frac{n_1 - n_2}{n_1 + n_2}, \quad r_3 = \frac{n_2 - n_3}{n_2 + n_3} \quad (4)$$

and the phase factor $\delta_i$ relates to the film thickness as:

$$\delta_1 = -i\left(\frac{2\pi}{\lambda}\right) n_1 d_1, \quad \delta_2 = -i\left(\frac{2\pi}{\lambda}\right) n_2 d_2 \quad (5)$$

The power transmission into Si is given by:

$$T = Re(n_3)|E_3^+/E_0^+|^2$$

$$= Re(n_3)\left|\frac{(1+r_1)(1+r_2)(1+r_3)\exp(\delta_1+\delta_2)}{1+r_1r_2\exp(2\delta_1)+r_1r_3\exp 2(\delta_1+\delta_2)+r_2r_3\exp(2\delta_2)}\right|^2 \quad (6)$$

where $E_0^+$, $E_3^+$, and $E_0^-$ are the amplitudes of the light waves incident, transmitted into the substrate, and reflected, respectively. Since we have assumed no absorption by $Al_2O_3$, the optical absorption (*A*) by the metal or graphene electrode becomes *A = 1 – T - R*. In Fig. 4 (a), we compare the absorption by a 10-nm-Au or single-layer-graphene electrode and the corresponding absorption by Si. A striking difference is that the absorption by 10-nm Au is more than 50% for photons with an energy higher than 4.5 eV whereas that by graphene remains low (< 6%) over the entire spectral range. Consequently, higher optical power can reach and be absorbed by Si with an enhancement from 10% ($Abs(Si)_{gra}$) to 40% ($Abs(Si)_{Au}$) in the high photon energy region, which is particularly important for extracting large energy barrier heights such as the energy barrier of holes. It is also worth noting that absorption by graphene depends on the oxide thickness. The oxide thickness used in IPE measurements is typically on the order of 10 nm, thus its optical thickness can be less than one tenth of the wavelength of a 6 eV photon. In test structures with thicker oxides, cavity enhanced absorption by graphene can be up to 4x higher.[24, 25] Fig. 4 (b) shows the improvement in semiconductor absorption by adopting the graphene transparent electrode: the ratio of absorption by Si to that by graphene is over one order of magnitude larger than the ratio of absorption by Si to that by Au for nearly the entire spectral range. This analysis supports our initial expectation that the graphene IPE electrode can substantially augment absorption by semiconductor while suppressing it by the collector electrode. As a result, the photocurrents are predominantly due to carrier injection from the semiconductor, which has led us to directly observe hole injection in an IPE measurement.

In summary, we have demonstrated a unique and experimentally facile approach to extract a complete energy band alignment using IPE measurements by employing graphene as the semitransparent collector electrode. Though the demonstration was performed on a model structure using $Al_2O_3$/Si, this technique can be broadly and readily extended to most structures studied by IPE measurements. The method presented here is largely free from experimental complexities and limitations commonly experienced by the prior methods, thus representing a milestone in the advancement of IPE metrology and the field of semiconductor interface studies.

Acknowledgements

The authors gratefully acknowledge the support of the NIST Semiconductor and Dimensional Metrology Division and the Nanoelectronics Research Initiative through the Midwest Institute for Nanoelectronics Discovery (MIND).

**Corresponding Author:**

Huili G. Xing: hxing@nd.edu;

Nhan. V. Nguyen: nhan.nguyen@nist.gov.

**Figure captions:**

**Figure 1.** (a) Schematic illustration of the graphene-$Al_2O_3$-Si device structure. Light illuminates from the top at normal incidence. (b) Optical image of the fabricated device. (c) Measured photocurrent as a function of incident photon energy. Gate voltage is applied to modulate the electric field in the oxide. (d) and (e) Schematic illustrations of electron and hole transitions determined by the direction of the oxide electric field.

**Figure 2.** (a) and (b) Cubic root of the quantum yield obtained by normalizing photocurrent to the incident light flux. The threshold of the yield varies with the applied gate voltage.

**Figure 3.** Schottky plots of electron and hole carrier injections as a function of the square root of the electric field. The linear extrapolation to zero field gives rise to the zero-field barrier height.

**Figure 4.** (a) Modeled optical absorption by graphene (Abs(Gra)), 10 nm Au (Abs(Au)), and Si (Abs(Si)$_{gra}$ and Abs(Si)$_{Au}$). (b) The ratio of graphene absorption over Si in a graphene-$Al_2O_3$-Si structure and that of Au absorption over Si in an Au-$Al_2O_3$-Si structure.